\documentclass[a4paper]{toledoStyle}
\usepackage{epsfig}

\begin{document}

\title{FIRST Extra-Galactic Surveys: Practical Considerations}

\author{S. J. Oliver} 

\institute{
Astronomy Centre,
University of Sussex, 
Brighton,
BN1 9QJ,
U.K., S.Oliver@Sussex.ac.uk}

\maketitle 

\begin{abstract}

We discuss some of the practical considerations that need to be made
in the design of extra-galactic field surveys using FIRST.  We
investigate the various limitations that confusion noise imposes on
possible FIRST surveys and the benefits of super resolution or decconvolution
techniques.  We consider the possible sizes and depths of survey
fields in order to meet many of the scientific objectives of the
FIRST mission.  In particular we discuss the factors that need to be
taken into account in selecting the survey fields.  The final choice
of fields will be influenced by the location of other surveys that
are currently being planned; so it is vital that the FIRST community
begin debating this issue now.  We conclude that a substantial survey
of order of 100 square degrees is likely to be sited in the region of
the SIRTF Legacy survey, SWIRE.

\keywords{Galaxies: formation -- Galaxies: surveys -- Missions: FIRST/Herschel}
\end{abstract}

\section{Choices in Designing a Survey}

A primary objective of the FIRST mission is to explore infrared loud populations in the distant Universe. As discussed by other speakers at this meeting it is anticipated that in order to meet this objective FIRST will spend a substantial fraction of its life undertaking (and following up) extra-galactic field surveys.  Such an investment requires careful preparation well in advance.    
The key parameters about which decisions have to be made in the planning of a survey are obvious enough; the wavelength, the depth, the total area and the specific fields.  The wavelengths of the FIRST filters have already been broadly defined and for SPIRE all filters operate simultaneously, so there is little to be decided there so we will discuss the three other factors in turn.

\section{Depth}
The maximum depth of FIRST surveys are expected to be limited by confusion so we will briefly review the classical confusion arguments and what they suggest for the limit to FIRST observations before discussing what gains might be made by employing super-resolution techniques.

\subsection{Classical Source Confusion}
With finite spatial resolution it is possible to identify bright sparse populations, but numerous faint populations become blurred and produce a background {\em confusion noise}.  This noise has been extensively discussed in the literature and we borrow heavily from a classic discussion by Condon (1974).
The distribution of intensity in a image, usually refereed to as  $P(D)$ distribution can be characterised by a ``noise'' 

\begin{equation}
\sigma_{\rm src} = \int_0^{S_{\rm lim}} S^2 dn
\end{equation}

where $S_{\rm lim}$ is a threshold, $S$ is the flux 
and dn is the number of pixels with that flux. 
The threshold, $S_{\rm lim}$, (necessary for a finite integral and to which $\sigma_{\rm src}$ is sensitive) gives an indication of 
the level above which sources can be identified; we can set $S_{\rm lim}=q\sigma_{\rm src}$, with e.g. $q=3,5$ depending on our bravado. 
A careful estimate of the source confusion noise would explicitly calculate the integral in this equation.
However, assuming power-law counts 
\begin{equation}\frac{dN}{dS}=-kS^{-\gamma},\end{equation}
with $\gamma= 5/2$ for Euclidean Counts, Condon (1974) showed that the

\begin{equation} \sigma_{\rm src}=\left(
\frac {q^{3-\gamma}} {3-\gamma} \right) ^ {1/(\gamma-1)}
(k\Omega_{\rm eff})^{1/(\gamma-1)},\end{equation}\label{eqn:sigma}
with the effective beam
\begin{equation}\Omega\_{\rm eff}=\int[f(\theta,\phi)]^{\gamma-1}d\Omega.\end{equation}
It is easily shown that the number density of sources at our confusion limiting flux ($S_{\rm lim}=q\sigma_{\rm src}$) is 
\begin{equation}n_q=\frac{1}{q^2}\frac{3-\gamma}{\gamma-1}\Omega_{\rm eff}^{-1},\end{equation}\label{eqn:nconf}
or
\begin{equation}n_q=\frac{1}{q^2}\Omega_{\rm eff}^{-1},\end{equation}
if $\gamma=5/2$.
It is this expression that leads to various ``rules of thumb'' for the number density at the source confusion 
Limit.
[For a Guassian beam the effective beam 
\begin{equation}\Omega_{\rm eff}=\frac{1}{(\gamma-1) \ln 2}\frac{1}{4}\pi\theta_1\theta_2\end{equation}
where $\theta_1,\theta_2$ are the FWHM of the beam, if $\gamma=5/2$, for an Airy beam we find 
\begin{equation}\Omega_{\rm eff}=0.18\pi\left(1.2\frac{\lambda}{D}\right)^2.\end{equation}  

To demonstrate that this classic confusion noise estimate is a useful definition of the survey limit we examine some real surveys.  With ISO at 15$\mu m$, $\Omega_{\rm eff}= 6\times 10^{-3}$ arcmin$^2$, thus the 5$\sigma_{\rm src}$ source density confusion limit would be $n_5=2.2$ arcmin$^{-2}$, or 43 sources in 
a 2.5$^\prime$ radius circle, c.f. Oliver et al. 2001, who reliably detect around 30 sources in that area.  Similarly the classical 5$\sigma_{\rm src}$ confusion limit for SCUBA at 850$\mu$m is 0.43 sources arc min$^{-2}$
 i.e. 3.8 sources in an area of
 8.7 arc min$^2$  c.f. 5 sources detected by 
Hughes et al. 1998

\subsection{Classical Confusion Limits from FIRST}
Applying this rule of thumb to his number count models Rowan-Robinson
2000\nocite{RR2000} estimated the confusion limits in Table \ref{tab:conf}.

\begin{table}
\caption{Source confusion limits estimated using the count models of Rowan-Robinson 2001}\label{tab:conf}
\begin{tabular}{lrrrrrr}
$\lambda \mu$m & 70 &120 & 175 & 250 & 350 & 500\\
$\Omega_{\rm eff}$/arcsec$^2$&13.9&40.7&86.6&176.8&346.4& 707.0 \\
$n_5$ & 12469 & 4243 & 1995 & 978 & 499 & 244\\
$4.3\sigma_{\rm src}/$mJy&0.74&3.2&11&18.6&20&16.6\\
\end{tabular}
\end{table}

Relatively short integrations with FIRST would reduce all other sources of noise to below these levels.

\subsection{Limits to Super Resolution}

It is tempting to enquire if super-resolution techniques will enable us to overcome these classical limits.  With no noise and a fully specified PSF we could deconvolve any image of point sources to recover the positions and fluxes of the sources. In practice significant deconvolution requires extravagant signal-to-noise ratios and places stringent demands on the determination of the PSF.   Lucy (1991, 1992a, 1992b)\nocite{lucy1991}\nocite{lucy1992}\nocite{lucy1992b}
investigated the theoretical limitations of super-resolution for two purposes. Firstly he discussed the descrimination between an extended source and an isolated point source. Assuming a perfect instrument which recorded the position of every photon, he demonstrated that the improvement in resolving power offered by any super-resolution technique as a function of number of photons $N$ was fundamentally constrained to
\begin{equation}\frac{\theta_{\rm Natural}}{\theta_{\rm Super}}<N^{1/4},\end{equation}
since it depends on the second moment of the image (Lucy 1991\nocite{lucy1991}).
In his second analysis he considered the discrimination between two close point sources and an elongated extended object and demonstrated that the
improvements were limited to
\begin{equation}\frac{\theta_{\rm Natural}}{\theta_{\rm Super}}<N^{1/8},\end{equation}
as this depends on the fourth moments (Lucy 1992a, 1992b)\nocite{lucy1992}\nocite{lucy1992b}. (Lucy also used numerical simulations to demonstrate that the constant of proportionality was close to unity, i.e. the techniques did not offer much benefit unless the signal-to-noise was high.

The confusion noise issue does not directly correspond to either of these two simple cases, 
It does however seem highly implausible that super-resolution techniques could offer better improvements than the first example and more likely that they would offer improvements comparable to the second example.  Thus we would expect the effective beam to improve 
$\Omega_{\rm eff}\propto\theta^2\propto N^{-\eta}\propto t^{-\eta}$, with $1/16<\eta<1/8$, i.e. from equation \ref{eqn:sigma}
$\sigma_{\rm src}\propto \Omega_{\rm eff}^{1/(\gamma-1)}\propto t^{-\eta/(\gamma-1)}$ or
a confusion limiting source density of 
$ n_q \propto\Omega_{\rm eff}^{-1} \propto t^\eta,$ (equation \ref{eqn:nconf})

Thus an increase in integration time from 15 minutes to 100 hours would allow you to 
increase the detectable number density of sources by a factor between 1.5 and 2, i.e. reducing the 
confusion noise limit by a factor of between 1.3 and 1.6.  

It thus seems unwise to integrate much longer than is required to reach the classical source confusion limit as any benefit that this imparts to the
super-resolution will be slight (and probably eradicated when one considers the uncertainties in the PSF).  We have not discussed the normalisation of these scaling relations, and so 
while we see that there is limited  value in undertaking  longer integrations, 
we have not assesed  the overall benefit of super resolution techniques.

\section{Area}

We have argued that there is not much benefit in surveys deeper than that
required to reach the classical confusion limit; since it doesn't
take long to reach this limit,  there is little point in small
area surveys.  As Planck will be undertaking an all sky survey
covering some of the FIRST wavelengths, it is clear that the most
productive niche for FIRST will be to undertake surveys of order 100
square degrees.  Other speakers show that these surveys are necessary
and sufficient to address the scientific issues of detecting
statistically significant samples of galaxies and to detect rare
high-luminosity objects at high redshift.

\section{Fields}
There are a number of factors that could influence the choice of
Survey field.  Primarily the fields should be chosen to the advantage of the FIRST data.  Factors which could influence this include, the zodical background,
The cirrus background, and visibility of the fields to the satellite.
\subsection{Factors Affecting Choice of Field}

\subsection{Zodiacal Background}
The zodical background is not expected to be a major problem, except perhaps at very low ecliptic latitudes and short wavelengths, such ecliptic latitudes would be excluded by our visibility constraints discussed next.

\subsection{Visibility}
Since the surveys are expected to be a dominant part of the mission life we want the fields to be visible for a substantial fraction of the time.  The longer their visibility, the less the survey programmes will impact on other targeted programmes, the easier it will be to follow-up survey sources with FIRST and the more scope there will be to design a favourable survey geometry. The visibility constraints are that the solar elongation must be between 60 and 120$^\circ$; averaged over a whole year this is simply a function of ecliptic latitude.  We suggest that a visibility of 50\% would be minimum, this corresponds to an ecliptic latitude of $|\beta|>45^\circ$. 
This criteria could be relaxed if there are more fields scattered across the sky.

\subsection{Cirrus Confusion}
Confusion can arise not just from faint blended sources, but from dust clouds within our own galaxy.  To estimate the level of this we use a scaling relation ship derived by Helou et al (1991), derived in turn from the power spectrum analysis of cirrus clouds performed by Gautier et al (1992).  
\begin{equation}\frac{\sigma_{\rm cir}}{\rm 1mJy}\approx
\left(\frac{\lambda}{\rm 100\mu m}\right)^{2.5}
\left(\frac{D}{\rm m}\right)^{-2.5}
\left(\frac{B(\lambda)}{\rm 1MJysr^{-1}}\right)^{1.5}.
\end{equation}
This relates the cirrus noise to the mean cirrus background intensity.  Relating this noise to the 
limiting source confusion noise
via $20\sigma_{\rm cir}=4.3\sigma_{\rm src}$ allows us to estimate that maximum safe 100$\mu$m intensity $B_{100}$.  The factor of $\sim 5$ incorporated here is a 
healthy safety margin.  The power spectrum analysis of the ISO 
175$\mu$m data in the Marano field demonstrated that in that field it
was possible to distinguish between real sources and cirrus fluctuations 
(\cite{lagach}), examination of their results arguably suggest that a less favourable field would
not have been able to do this. 
Our scaling analysis in this fields suggests,
4.3$\sigma_{\rm src}=107$ mJy, and taking $B_{100}$ = 0.88 MJy/sr
10$\sigma_{\rm cir}= 116$ mJy,  for comparison  the source lists which were  extracted to 100 mJy
sources extracted to 100 mJy

The limit imposed by this conservative constraint is $B_{100} < 2$ MJy/sr.
A similar analysis for SIRTF suggests that the constraint is $B_{100} < 2$ MJy/sr

\begin{figure*}
\epsfig{file= 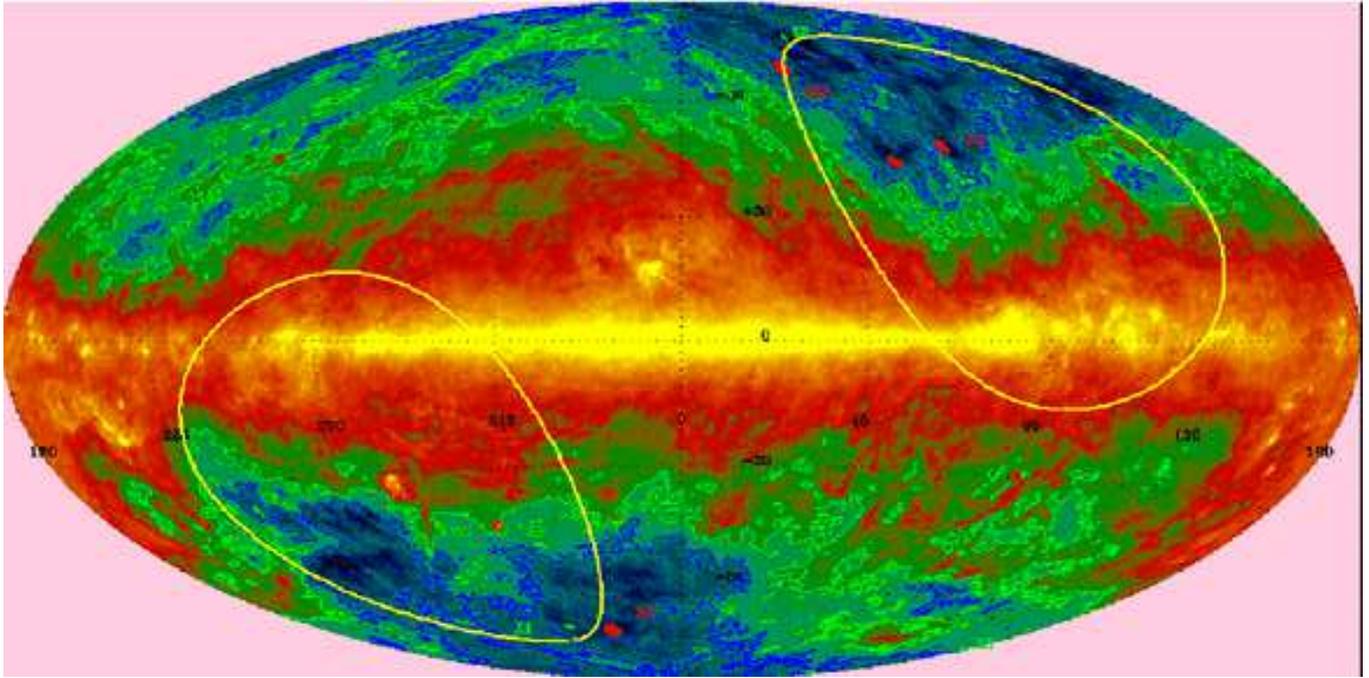, angle =270, width=18cm}
\caption{Mean 100 $\mu$m intensity from the maps of Schlegel et al. (1998), plotted in
galactic coordinates.  Overplotted are contours at 1 (blue) and 2 (green) MJy/sr, the limits
suggested for SIRTF and FIRST surveys respectively.  Also plotted is a line at $|\beta |=45^\circ$.}\label{fig:i100}
\end{figure*}

The cirrus confusion constraints and the suggested ecliptic latitude constraints are
illustrated in figure \ref{fig:i100}

\subsection{Existing Survey Data}

Good supporting data at other wavelengths is expected to be critical
to realising the scientific goals of FIRST surveys.  The FIR/Sub-mm
SEDs of the sources are basically black bodies and the degeneracy
between temperature and redshift means that the FIRST data alone will
poorly constrain the nature of the sources. Since the sources are
expected to be at high redshifts (and dusty) the optical counterparts
are expected to be very faint. As has been found with SCUBA
observations, a comparatively large beam, exacerbated by source
confusion noise leads to an error circle large compared to the number
density of possible counterparts.  This will makes identification
extremely difficult.  The more complimentary data there is at other
wavelengths the more the identification process can be refined.  
Wavelengths which probe emission more tightly correlated with the 
FIRST emission will be particularly useful.  Radio and infrared
surveys are likely to be particularly suitable. 
One survey of particular note is the 
SIRTF Wide-area InfraRed Extra-galactic survey (SWIRE, PI Carol Lonsdale
\verb+http://www.ipac.caltech.edu/SWIRE/+).
Almost identical considerations motivated the choice of survey fields for this
project, (in fact the cirrus constraints were even more stringent than necessary
for FIRST).  As this survey will cover 70 square degees and will have a vast wealth of complemntary
data at all wavelengths by the time FIRST flies, it will be an excellent target for
the FIRST surveys.

\section{Conclusions}

We have considered various aspects of the design for surveys with FIRST, employing simple scaling relationships which are justified by comparison wth existing ISO and SCUBA surveys.
We have argued that integration significantly below the classical confusion limit will not provide any substantial benefit to super-resolution techniques.  FIRST's niche is thus expected to be for surveys of order 100 square
degrees, with Planck exploring the parameter space of larger area, shallower surveys.  Cirrus constrains these surveys to be at $B_{100}<2 $MJy/sr and visibility considerations prefer $| \beta |>45^\circ$.  Distributing number of survey fields would aid visibility and ground based follow-up.  We recognise the very high importance of complimentary data at other wavelengths and note that the restrictions on the FIRST survey fields are more than adequately met by the fields selected for the SIRTF Legacy programme SWIRE.

\begin{acknowledgements}
Thanks to the organisers for an enjoyable conference.  Thanks in particular to Goran and the other editors for their almost limitless patience!

\end{acknowledgements}

\end{document}